\documentclass[12pt]{article}
\usepackage{epsf}
\usepackage{graphicx}
\usepackage{amssymb}
\usepackage{epstopdf}
\usepackage{amsmath}
\usepackage{amssymb}
\usepackage{epsfig}
\usepackage{cite}
\usepackage{color,colordvi}
\usepackage{graphicx}
\usepackage{pdfpages}
\graphicspath{ {./} }
\usepackage[left=2.5cm,bottom=3cm,right=2.5cm,top=3cm]{geometry}

\usepackage[latin,english]{babel}
\usepackage{amsmath}
\usepackage{amssymb}
\usepackage{graphicx}
\usepackage{mathtools}
\usepackage{tikz} 
\usepackage{epsf}
\usepackage{graphicx}
\usepackage{amssymb}
\usepackage{epstopdf}
\usepackage{amsmath}
\usepackage{amssymb}
\usepackage{epsfig}
\usepackage{cite}
\usepackage{color,colordvi}
\usepackage{graphicx}
\usepackage{pdfpages}
\graphicspath{ {./} }
\usepackage[left=2.5cm,bottom=3cm,right=2.5cm,top=3cm]{geometry}

\usetikzlibrary{patterns}
\usetikzlibrary{decorations.markings,decorations.pathmorphing}
\usetikzlibrary{calc}
\tikzstyle{singularity}=[line width=0.6,decorate,
                         decoration={zigzag,amplitude=2,segment length=6.17}]
\usetikzlibrary{shapes.misc}
\colorlet{mydarkred}{red!50!black}
\begin{document}
\vspace{0.01cm}
\begin{center}
{\Large\bf  Inflationary Cosmology as flow of integrable weights} 

\end{center}

\vspace{0.1cm}

\begin{center}

{\bf Cesar Gomez}$^{a}$\footnote{cesar.gomez@uam.es}

\vspace{.6truecm}

{\em $^a$
Instituto de F\'{\i}sica Te\'orica UAM-CSIC\\
Universidad Aut\'onoma de Madrid,
Cantoblanco, 28049 Madrid, Spain}\\

\end{center}

\begin{abstract}
\noindent  
{ We identify the algebra of gauge invariant observables in de Sitter as the subalgebra of the type $III_1$ factor $A_{dS}$ associated to de Sitter, defined as the centralizer of any integrable weight on $A_{dS}$. These algebras are for any integrable weight type $II_{\infty}$ factors admitting a crossed product representation with respect to modular automorphisms. In this context we define Inflationary Cosmology as the {\it flow of integrable weights} and the {\it dual automorphism} as the flow generator. Using some basic properties of integrable weights we show that any type $II_1$ dS algebra {\it cannot be} represented as the $\epsilon=0$ limit ( for $\epsilon$ the slow roll parameter ) of the gauge invariant algebra defined by any integrable weight. This strongly indicates that the pure dS algebra defined as the $\epsilon=0$ limit is algebraically non existent.

}

\end{abstract}

\thispagestyle{empty}
\clearpage
\tableofcontents
\newpage
\section{Preliminary Comment}
The following note is written as a contribution to the forthcoming book in the memory of Valeria Rubakov. In the summer of 1986 I met Valeria in the Th division at Cern. John Ellis introduced me to Valeria at her 40th birthday celebration. A spirit of friendship was immediately created and we began to discuss physics already during the party. During the same period Sander Bais was also visiting Cern and we started a nice and very alive collaboration on something that at that time was a bit obscure. We considered {\it cobordism} to identify what global charges can be conserved in presence of quantum gravity. These discussions materialize in a join paper \cite{Rubakov}. Many years later, actually quite recently, the idea of cobordism has received a new {\it quantum} impulse in the swampland framework \cite{vafa}. The present note written in his memory is not directly related to our discussion on cobordism but is certainly very much related to the sort of quantum gravity ideas that occupied Valeria's mind throughout her unfortunately too short life.

\section{Introduction}
The recent {\it algebraic program} to quantum gravity is based on mapping bounded regions of space into von Neumann algebras acting on a separable Hilbert space. These von Neumann algebras are assumed to be type $III_1$ factors and to represent the algebra of QFT local observables in the weak gravity limit. The use of type $III_1$ factors to define the algebras of observables, in the weak gravity limit $G\sim 0$, of bounded regions of space began in a series of important programatic papers \cite{L},\cite{LL1},\cite{LL2},\cite{Witten1} among others. The two basic physics targets of this algebraic program, from the limited point of view of the present note, have been the identification of the {\it gauge invariant algebra of observables} and the identification and precise definition of {\it generalized entropies} (see among others \cite{Witten2},\cite{Witten3},\cite{K1},\cite{K2},\cite{Gomez3}). In this note we will reduce our discussion to the problem of defining the gauge invariant algebra mostly focusing on {\it inflationary cosmological backgrounds}.

In order to get immediately in contact with the target of this note let us focus on the algebra of observables, in the weak gravity limit, associated to a generic de Sitter observer worldline. Following \cite{Witten2} we will assume that this algebra is a type $III_1$ factor that we will denote $A_{dS}$ ( see next section for some basic definitions ). The problem on which we will concentrate our attention is on the definition of the corresponding {\it gauge invariant subalgebra} that we will denote temporarily as $A_{dS}^{phys}$. In GR and for the particular case of de Sitter this algebra is defined as the set of observabls in $A_{dS}$ {\it invariant} under coordinate time reparametrizations. Algebraically these time reparametrizations are defined by the {\it modular automorphisms} associated to a generic {\it state} $\phi$ on $A_{dS}$
\footnote{We will assume this state is normal and faithful.}. As observed in \cite{Witten2} the so defined gauge invariant algebra is trivial and equal to multiples of the identity (see section 2.3 for an explanation). In order to improve this situation and to define a rich enough gauge invariant subalgebra we need, as we will discuss in the ext section, to relax the condition of $\phi$ to be a {\it state} and to work with {\it weights} on $A_{dS}$ in particular with {\it integrable weights} \cite{CT}.

A popular approach, introduced in \cite{Witten2}, to the physics meaning of these integrable weights is to interpret them as representing, in addition to the quantum field theory fluctuations localized in a dS static patch, {\it an extra external quantum observer equipped with some clock}, where by that we just mean any way to assign time to the observer measurements ( we will discuss the algebraic interpretation of these observers in detail in sections 2 and 3).

At this point we can summarize the definition of gauge invariant subalgebra $A_{dS}^{phys}$ as the {\it centralizer} $(A_{dS})_{\omega}$ i.e. the subalgebra of $A_{dS}$ invariant under the modular automorphisms $\sigma_{\omega}^t$ defined, in the Tomita-Takesaki formalism, by an integrable weight $\omega$. This gauge invariant algebra has some physically relevant properties. Next we enumerate some basic ones:
\begin{itemize}
\item i) For {\it any} integrable weight $\omega$ we can find an {\it isomorphism} $A_{dS}=P\otimes F_{\infty}$ with $P$ isomorphic to $A_{dS}$ and $F_{\infty}$ the algebra of bounded operators on $L^2(\mathbb{R})$ i.e. the algebra of {\it quantum observables} of an elementary quantum mechanical system with Hilbert space $L^2({\mathbb{R}})$ \footnote{This factor will be the natural one to associate with the {\it quantum external observer} of \cite{Witten2}.}. Relative to this isomorphism $\omega$ becomes $\phi\otimes \bar \omega$ for $\phi$ a weight on $P$ and $\bar \omega$ the weight on $F_{\infty}$ defined by \cite{CT} $\bar\omega(x)=tr(Ux)$ for $x$ any element of $F_{\infty}$ and $U$ the unitary such that $U^{it}$ generates the translations on $L^2(\mathbb{R})$.
\item ii) For any integrable weight $\omega$ the gauge invariant algebra $(A_dS)_{\omega}$ is a type $II_{\infty}$ factor that can be represented as {\it the crossed product} $A_{dS}\rtimes_{\sigma_{\phi}^t} {\mathbb{R}}$ for $\phi$ the weight introduced in item i) above\footnote{As we will discuss this crossed product representation for {\it any} integrable weight follows from the important result \cite{CT} that any integrable weight on $A_{dS}$ is {\it dominant}.}.
\item iii) For any integrable weight $\omega$ the gauge invariant algebra $(A_{dS})_{\omega}$ is equipped with a trace $Tr_{\omega}$ \footnote{This follows from the KMS property for $\omega$.}. 
\item iv) Relative to $Tr_{\omega}$ we can define {\it affiliated density matrices} that will be relevant in the discussion of generalized entropies. We will not touch this discussion in this note.
\item v) On the gauge invariant algebra we can define the {\it dual} to the modular automorphisms $\sigma_{\phi}^t$. We will denote these automorphisms $\theta_{\phi}^s$ with $s$ representing the {\it dual modular coordinate}. This fact will be the most important for our discussion in this note on the definition of Cosmology as flow of weights. 
\end{itemize}

Let us now say few words on the algebraic approach to Cosmology, in particular to {\it Inflationary Cosmology}. Attempts to identify the algebra of gauge invariant observables in the background of Inflationary Cosmology have been presented previously in \cite{Gomez1},\cite{Gomez2}, and more recently in \cite{K} and \cite{Penington}. 

In the context of Cosmology the identification of the gauge invariant algebra is more subtle due to the {\it cosmological} time dependence. Indeed in Inflationary  Cosmology we expect to define physical quantities, invariant under {\it coordinate reparametrizations}, depending non trivially on the cosmological time. This cosmological time evolution should reflect the cosmological evolution of the Universe in particular, in the case of Inflation, the graceful exit. In Inflation this time dependence of gauge invariant observables will be derived from the {\it Einstein equations} for some {\it inflaton potential}. The question we would like to face, in order to define some algebraic version of Cosmology, is: {\it How to account, in the algebraic setup, for the cosmological time dependence of gauge invariant observables?}

Once, we define invariance under coordinate reparametrizations using the centralizer of some {\it integrable weight}, the simplest possibility to define {\it cosmological time evolution}, in the inflationary context, is to consider one parameter families, let us say $\omega_s$, of integrable weights on $A_{dS}$ with $s$ parametrizing the cosmological time. In this context the cosmological time dependence of observable quantities is defined by the $s$ dependence of $\omega_s(\hat a)$ for $\hat a$ elements in $(A_{dS})_{\omega_s}$. Thus the question becomes: How to identify the $s$ dependence representing the flow of integrable weights ? 

Here the problem is simplified as a consequence of the type $III_1$ nature of $A_{dS}$. Indeed, in this case the {\it flow in $s$ of integrable weights is essentially trivial and reduces to weight re-scalings.} \cite{CT}. In summary the algebraic approach to (Inflationary) Cosmology we will develop in this note is based on \footnote{The role of dual automorphisms for the case of the black hole time dependence during the quantum  evaporation has been recently considered in \cite{Gomez-dual}. We will not discuss this problem in this note.}:

\vspace{0.3 cm}

{\it To model Cosmological time evolution as flow of integrable weights on a type $III_1$ factor $A_{dS}$.}

\vspace{0.3cm}

Recently \cite{Penington} some concrete examples of integrable weights on $A_{dS}$ have been constructed on the basis of the Hartle-Hawking formalism fo some simple euclidean action for the inflaton field \footnote{See also \cite{Witten4},\cite{Malda}.}. We will briefly discuss the meaning of these weights as examples of {\it dominant weights} as well as the possible interpretation of the Cosmological time evolution as {\it flows of these integrable weights}.

In Inflationary Cosmology we are familiar with the construction of {\it gauge invariant} observables representing, for instance, the primordial scalar fluctuations. These variables are known as Mukhanov-Sasaki variables ( \cite{MCH1}\cite{MCH2} \cite{MSteinhardt} \cite{Sasaki}). These variables are defined in Inflationary backgrounds characterized by a non vanishing value of the slow roll parameter $\epsilon$ defined by the equation of state associated to the inflaton potential. In addition these variables depend on the cosmological time according to the Mukhanov-Chibisov \cite{MCH1} equation of motion. In the algebraic setup, that we suggest in this note, these variables should be formally defined as continuous maps from the {\it flow coordinate $s$}, that parametrizes the particular flow of integrable weights as $\omega_{s}$, into elements in the different centralizers $(A_{dS})_{\omega_{s}}$ associated to the flow. 
In this note we will reduce ourselves to some general comments for the simple flow of integrable weights on $A_{dS}$. 

An important issue that we will address in this general context is the problem of defining the algebra of gauge invariant observables in {\it the pure dS limit defined by $\epsilon=0$}. In this limit Mukhanov-Sasaki variables are ill defined as it is also the case for the power spectrum of scalar fluctuations \footnote{Both diverging as $\frac{1}{\epsilon}$.}. The obvious question is: How should we interpret this problem?

Using the integrable weight approach to the identification of the gauge invariant algebra of observables we should expect that the algebra of gauge invariant observables for pure dS defined in the limit $\epsilon=0$ should be the centralizer of some particular integrable weight, let us say $\tilde\omega$, describing pure dS {\it in the $\epsilon=0$ limit}\footnote{Note that in the standard approach to Inflationary Cosmology the gauge invariant observables are not defined in the $\epsilon=0$ limit.}. In \cite{Witten2} it has been defined as the pure dS gauge invariant algebra a type $II_1$ factor. This factor can always be represented as $p(A_{dS})_{\omega}p$ for $\omega$ any integrable weight on $A_{dS}$ and $p$ a finite projection i.e. with finite value of $Tr_{\omega}(p)$ \footnote{With different finite projections $p$ corresponding to different algebraic versions of pure dS with potentially different values of the cosmological constant.}. Thus the question becomes: 

\vspace{0.3cm}

{\it Can we find a weight $\tilde\omega$ on $A_{dS}$ such that the centralizer $(A_{dS})_{\tilde\omega}$ is isomorphic  to the type $II_1$ factor $p(A_{dS})_{\omega}p$ for some finite projection $p$ ?}

\vspace{0.3cm}

Note that only in that case we could identify the type $II_1$ factor {\it as describing an $\epsilon=0$ limit}. We will argue that such weight $\tilde\omega$ {\it does not exist}. In fact any weight defining the pure dS type $II_1$ algebra as the corresponding centralizer will be a dominant weight and therefore its centralizer will be a type $II_{\infty}$ factor ( see Lemma 2 in section 4 of this note). Consequently we will conclude that: \footnote{In section 4 we will briefly extend the former argument. The quantum inconsistency of the pure dS $\epsilon=0$ limit has been treated previously from different points of view in particular from the {\it corpuscular point of view} in \cite{Gia1,Gia2,Gia3}. The analysis in this note is independent of any corpuscular model and it is purely algebraic.}

\vspace{0.3cm}

{\it The pure dS limit $\epsilon=0$ is algebraically inconsistent}

\vspace{0.3cm}

One interesting output of the algebraic approach presented in this note is to link the structure of the factor $A_{dS}$ as $P\otimes F_{\infty}$ and the {\it quantumness of $F_{\infty}$} with the quantum inconsistency of the $\epsilon=0$ limit. Morally speaking, it looks that non trivial cosmological evolution, reflecting non vanishing $\epsilon$, is quantum mechanically mandatory.

\section{de Sitter von Neumann algebras}
\subsection{General setup}
Following the basic result by Araki \cite{Araki} (see also \cite{AW}) on the type of factors defined by local QFT in bounded regions of space-time we will {\it assume} that associated to the static patch of an ideal, but generic, dS observer wordline, we can define a type $III_1$ von Neumann factor $A_{dS}$ acting on a separable Hilbert space ${\cal{H}}$.
These data define the {\it commutant} $A^{'}_{dS}$ as the set of bounded operators in $B({\cal{H}})$ commuting with all elements in $A_{dS}$. We will assume that the center of $A_{dS}$ is 
${\mathbb{C}}1$ and consequently that $B({\cal{H}})=A_{dS}\otimes A^{'}_{dS}$. We will also assume that the factors associated to {\it different observers are  spatially isomorphic} \footnote{This means that the factors $A_{dS}$ associated with different observers i.e. with different static patches, are not only algebraically isomorphic but also the existence of an isomorphism between the different Hilbert spaces on which these factors are acting that maps the corresponding factors. The notion of spatial isomorphism is crucial in the classification problem of factors (see \cite{MvN4}).}. 

Physically the key consequence of the assumption about the type $III_1$ nature of $A_{dS}$ is the absence of a {\it factorization of the Hilbert space ${\cal{H}}=H_1\otimes H_2$} such that $A_{dS}=B(H_1)\otimes 1$ and $A^{'}_{dS}=1\otimes B(H_2)$. This can be interpreted, in case we think $A_{dS}$ as the algebra of local observables with support in a given static patch and $A^{'}_{dS}$ as the algebra of local observables on the {\it mirror patch}, as reflecting {\it infinite entanglement} between these patches. 

\subsection{States and centralizers} 
We will define a {\it state} $\phi$ on $A_{dS}$ as a {\it finite} linear form  that we will assume to be normal and faithful. We can use the linear form $\phi$ to equip $A_{dS}$ with a pre-Hilbert structure defined by the scalar product $\langle a,b\rangle=\phi(a^*b)$ with $a$ and $b$ generic elements of $A_{dS}$. We will denote this pre-Hilbert space ${\cal{H}}_{\phi}$. The natural {\it representation} of $A_{dS}$ as bounded operators acting on ${\cal{H}}_{\phi}$ defines, after appropriated completion, the GNS representation of $A_{dS}$ associated to $\phi$.

Tomita-Takesaki theory (see \cite{Jones} for a good review) allows us to associate to any {\it state} $\phi$ the one parameter group of {\it modular automorphisms} that we will denote $\sigma_{\phi}^t$ with $t$ the {\it modular time} group parameter. The most important property of the modular automorphism $\sigma_{\phi}^t$ is to identify the state $\phi$ as the {\it unique normal state} satisfying the KMS property with respect to the modular automorphism. 

From the KMS property follows a physically crucial fact, namely that the state $\phi$ will satisfies the {\it trace property} on the subalgebra of $A_{dS}$ defined by those elements $a\in A_{dS}$ {\it invariant} under $\sigma_{\phi}^t$ i.e. $\{ a\in A_{dS}; \sigma_{\phi}^t(a)=a\}$. This subalgebra  is known as the {\it centralizer} and will be denoted as $(A_{dS})_{\phi}$. 

The centralizer subalgebra is importat for two reasons. First of all it allows us to identify the subalgebra of $A_{dS}$ on which we can define a {\it trace} which physically means that we can assign average values to the observables in $(A_{dS})_{\phi}$. The second reason is that this algebra admits the interpretation of the {\it gauge invariant subalgebra} provided we interpret the modular automorphism $\sigma_{\phi}^t$ as defining {\it time reparametrizations}.

As stressed in \cite{Witten2} for $A_{dS}$ a type $III_1$ factor and for $\phi$ a {\it state} \footnote{From now on we will assume the state is normal and faithful.} the corresponding {\it centralizer} $(A_{dS})_{\phi}$ is {\it trivial} and isomorphic to the multiples of the identity. The reason is that in this case $\sigma_{\phi}^t$ is acting {\it ergodically} on $A_{dS}$. Thus for $\phi$ a {\it state} the algebra of gauge invariant observables reduces to just multiples of the identity. This fact was used in \cite{Witten2} to motivate the inclusion of an {\it external quantum observer} equipped with a {\it clock} that {\it says the "time" at which the corresponding measurement of any observable in $A_{dS}$ is taking place.}

As already noticed by Page and Wooters \cite{PageW} this problem is reminiscent of the general problem we find with superselection charges with now the Tomita Takesaki modular Hamiltonian $h_{\phi}$ playing the role of the superselection charge. The natural solution, inspired by the original discussion of Aharonov and Susskind \cite{AS} is, in modern language \cite{Refe}, to add a quantum reference frame or more generically an external quantum observer \footnote{For the relation between quantum reference frames and observers see \cite{Gomez1} \cite{Hoem},\cite{Leigh} and the very complete set of references therein.}. 

Although this line of thought is intuitively very appealing and deep the very notion of external quantum observer is very vague and requires several clarifications. Hence we will start addressing the general problem of how, for type $III_1$ factors, we can define non trivial candidates for representing the algebra of gauge invariant observables. 

\subsection{Obervers versus Weights}
An important generalization of the notion of {\it state} is the notion of {\it weight}. In few words weights differ from states in that they take values in $[0,+\infty]$ and are defined on the positive part of the algebra $A_{dS}^+$. The  key property of weights, in the particular context of type $III_1$ factors, is that while {\it for  a normal state $\phi$} the {\it centralizer} $(A_{dS})_{\phi}$ is trivial, due to the {\it ergodicity} of the corresponding modular automorphism $\sigma_{\phi}^t$, this is not the case if instead of using the state $\phi$ we use an appropriated {\it weight} $\omega$ on $A_{dS}$ and the corresponding modular automorphism $\sigma_{\omega}^t$. In this case, as proved in Connes thesis \cite{Connes} for type $III_{\lambda}$ factors for $\lambda \in]0,1[$ and by Connes and Takesaki in \cite{CT} for type $III_1$ factors, we can find a particular type of weight $\omega$ such that, for the type $III_1$ i.e for the case of $A_{dS}$,:
\begin{itemize}
\item i) The corresponding centralizer $(A_{dS})_{\omega}$ is an infinite type $II_{\infty}$ factor
\item ii) The restriction of $\omega$ to $(A_{dS})_{\omega}$ defines a semi finite trace
\end{itemize}
Note that condition ii) above follows from the KMS property for the weight $\omega$.

The obvious question at this point should be: What is the relation between the weight $\omega$ satisfying conditions i) and ii) above and the "existence" of and {\it external quantum observer} ?

\subsection{The importance of being type $III_1$}
A reasonable physics question is in what sense the "observer" should be considered as an {\it external physical system} that should be distinguished from the typical degrees of freedom describing the gravitational, as well as other Q.F.T fluctuations, localized in the static patch. {\it This very intuitive question has not any {\it unique} answer in the case of type $III_1$ factors.}

The reason for that is the {\it properly infinite} nature of type $III_1$ factors \cite{CT}.

More specifically, for $A_{dS}$ a type $III_1$ factor, we can find different {\it representations} of $A_{dS}$ as 
\begin{equation}
A_{dS}=P\otimes F_{\infty}
\end{equation}
with $F_{\infty}$ the type $I$ factor defined by the algebra of bounded operators acting on $L^2({\mathbb{R}})$ and with $P$ {\it isomorphic} to $A_{dS}$ \footnote{Here by different representions of $A_{dS}$ as $P\otimes F_{\infty}$ we mean different isomorphisms.}

It could be tempting to interpret any of these isomorphisms of $A_{dS}$ as $P\otimes F_{\infty}$ as defining a {\it decomposition} of $A_{dS}$ into an algebra $P$ describing the localized quantum fluctuations ( gravitational as well as q.f.t) and to use the type $I$ factor $F_{\infty}$ to describe the {\it external quantum observer}.

Using this decomposition we can define a {\it weight} $\omega$ on $A_{dS}$ satisfying conditions i) and ii) above as
\begin{equation}\label{dominant}
\omega=\phi\otimes \bar \omega
\end{equation}
for $\phi$ some weight on $P$ and with $\bar\omega$ a weight on $F_{\infty}$
defined by \cite{CT}
\begin{equation}\label{observerweight}
\bar \omega (x)=tr(x U)
\end{equation}
for $x$ any element in $F_{\infty}$ and where $U^{it}$ is, for any $t$, the unitary representation of {\it translations} in $L^2({\mathbb{R}})$. 

We could interpret the so defined weight  $\omega$ on $A_{dS}$ as describing the quantum fluctuation dynamics in terms of $\phi$ and the {\it observer dynamics} i.e the external clock, in terms of $\bar \omega$ \footnote{This construction of $\omega$ was initially introduced in \cite{Connes} for type $III_{\lambda}$ with $\lambda\neq1$ as {\it generalized traces} (see theorem 4.2.6). For type $III_1$ the weight $\phi$ in the representation $\omega=\phi\otimes\bar\omega$ should be a {\it weight} on $P$ with infinite centralizer. Thus for type $III_1$ the weight $\phi$ cannot be a {\it normal state} on $P$ that will lead to a trivial centralizer.}.

\section{Integrability and gauge invariance}
As stressed we are interested in weights $\omega$ on $A_{dS}$ such that they satisfy conditions i) and ii) above i.e. they define a centralizer $(A_{dS})_{\omega}$ that it is a type $II_{\infty}$ factor. This implies that the algebra of gauge invariant observables is rich enough and that it is equipped with a semifinite trace.

\subsubsection{Integrability, dominant weights and gauge invariance}
The simplest, and physically more direct way to define observables which are gauge invariant with respect to the modular time reparametrizations defined by a generic faithful state $\phi$ on $A_{dS}$,  will be to define them as
\begin{equation}\label{integral}
\int_{-\infty}^{+\infty} dt \sigma_{\phi}^t(a)
\end{equation}
for $a$ any element in $A_{dS}$. It is easy to see that this integral is not defined for a generic element $a$ in $A_{dS}$ if we use a faithful {\it state} $\phi$. This is a different way to see that for any state $\phi$ on a type $III_1$ factor the centralizer $(A_{dS})_{\phi}$ is trivial and reduced to multiples of the identity. 

In order to define a non trivial centralizer for a weight $\omega$ we  need to look for weights for which
\begin{equation}\label{integral2}
\int_{-\infty}^{+\infty} dt \sigma_{\omega}^t(a)
\end{equation}
is well defined for a {\it dense} subset of $A_{dS}$. Weights satisfying this property are known as {\it integrable weights} \cite{CT}.

For an integrable weight $\omega$ we can define the map
\begin{equation}\label{projection}
E_{\omega}:A_{dS}\rightarrow (A_{dS})_{\omega}
\end{equation}
with
\begin{equation}
E_{\omega}(a)=\int_{-\infty}^{+\infty} dt \sigma_{\omega}^t(a)
\end{equation}
Obviously $E_{\omega}(a)$ for any $a\in A_{dS}$ is a gauge invariant observable. In this sense $E_{\omega}$ defines, in physics terminology, the {\it dressing of the observable $a\in A_{dS}$}. In \cite{Witten2} the element $E_{\omega}(a)$ defined for an integrable $\omega$ is normally denoted as $\hat a$ \footnote{Mathematically $E_{\omega}$ for an integrable weight $\omega$ defines a {\it conditional expectation} \cite{Takesaki1}. Note that $E_{\omega}$ is a projection i.e. $E_{\omega}^2=E_{\omega}$. For a discussion on gauge invariance and conditional expectations see also \cite{Leigh2} and references therein.}

In summary, {\it any integrable weight} $\omega$ defines a {\it gauge invariant subalgebra $(A_{dS})_{\omega}$}, namely the image of the projection $E_{\omega}$. Thus, in order to identify the {\it different {\bf physics} we can associate with a given patch} we need to identify and classify the full set of {\it integrable weights} on $A_{dS}$. Fortunately for the case of type $III_1$, that is the one relevant for $A_{dS}$, this job was done in \cite{CT}. Next we summarize some basic results.

\subsection{Dominant weights and crossed products}
For the type $III_1$ case i.e. for $A_{dS}$ the space of integrable weights is specially simple. In essence {\it for any integrable weight} $\omega$ it exits a representation of $A_{dS}$ as $P\otimes F_{\infty}$ with $P$ isomorphic to $A_{dS}$ and a weight $\phi$ on $P$ such that
\begin{equation}
\omega=\phi\otimes \bar\omega
\end{equation}
with $\bar \omega$ defined in (\ref{observerweight}). Intuitively we could interpret this result saying that any non trivial {\it gauge invariant physics} on a static patch can be described using a weight $\phi$ on $P$ to describe the quantum fluctuations and a weight $\bar\omega$ to describe the external observer.

The fact that for {\it any integrable weight} $\omega$ we can find a representation like $\omega=\phi\otimes\bar\omega$ for $\phi$ a weight on $P$ has a very interesting and deep consequence, namely:

\vspace{0.3cm}

{\bf Lemma: For any integrable weight the gauge invariant subalgebra $(A_{dS})_{\omega}$ can be represented as the crossed product
\begin{equation}\label{crossed}
(A_{dS})_{\omega} =P\rtimes_{\sigma_{\phi}^t} {\mathbb{R}}
\end{equation}
for $\sigma_{\phi}^t$ the modular automorphism defined by the weight $\phi$.}

\vspace{0.3cm}

This lemma follows from the basic result in \cite{CT} establishing that for type $III_1$ factors all integrable weights are dominant.

This crossed product representation is crucial to prove {\it independence} ({\it up to unitarity equivalence}) of the gauge invariant algebra $(A_{dS})_{\omega}$ on the weight $\phi$ on $P$ entering into the representation of $\omega$ as $\omega=\phi\otimes\bar\omega$ \footnote{It is tempting to interpret this independence on $\phi$ as {\it background independence}. Note that this independence follows from the crossed product representation. Indeed the crosed product (\ref{crossed}) is independent, up to unitary equivalence, on the weight $\phi$.}.

{\bf Remark} Recently in \cite{Penington} it has been suggested that once we include a cosmological inflaton and a non trivial slow roll potential $V^{'}\neq 0$ we can define an integrable weight, let us say $\omega_{cos}$ with a type $II_{\infty}$ centralizer but with not {\it crossed product representation}. As discussed in this section this is not possible if $\omega_{cos}$ is an integrable weight on a type $III_1$ factor. In a future section we will discuss the algebraic approach to slow roll inflationary Cosmology to put this technical comment in context.

\subsection{Emergent Duality}
Once we have proved that the gauge invariant subalgebra $(A_{dS})_{\omega}$ for {\it any integrable weight $\omega$} admits the crossed product representation
\begin{equation}\label{crossed}
(A_{dS})_{\omega} =P\rtimes_{\sigma_{\phi}^t} {\mathbb{R}}
\end{equation}
we can define on $(A_{dS})_{\omega}$ the {\it dual} to the modular automorphism $\sigma_{\phi}^t$. While the modular automorphism represents the action of the group $G$ of {\it time translations} on $A_{dS}$ the {\it dual} action defines the action of the dual group $G^*$ on the gauge invariant algebra $(A_{dS})_{\omega}$. The group $G$ of {\it time translations} is simply ${\mathbb{R}}$ and the dual is just the dual real line $R^{*}$ that we will parametrize with $s$. Thus the automorphism defining the action, {\it dual} to $\sigma_{\phi}^t$, on $(A_{dS})_{\omega}$ will be denoted $\theta_{\phi}^s$ \cite{CT}.

This dual action was originally introduced by Takesaki \cite{Takesaki}. The way this dual automorphism acts on gauge invariant observables is \cite{CT},\cite{Takesaki} given by
\begin{equation}\label{dual}
\theta_{\phi}^s(\hat a)= (1\otimes V_s)\hat a (1\otimes V_s)^{*} 
\end{equation}
for $\hat a$ any element in $(A_{dS})_{\omega}$ i.e. any gauge invariant observable 
$\hat a =\int_{-{\infty}}^{+\infty} \sigma_{\omega}^t(a)$.

The operator $V_s$ is the Weyl {\it dual} in $F_{\infty}$ to the generator $U_t$ defining the action of translations on $L^2({\mathbb{R}})$. Note that the reason for using the term {\it duality} are the Weyl relation representing the Heisenberg algebra between $U_t$ and $V_s$ namely
\begin{equation}\label{Weyl}
V_sU_t= e^{ist}U_tV_s
\end{equation}

\vspace{0.5cm}

{\bf Remark}
Let us pause a moment to recall some basic properties of the type $I$ factor $F_{\infty}$. This factor is just the algebra of bounded operators acting on $L^2({\mathbb{R}})$ i.e. the algebra of observables of an elementary quantum mechanical system with infinite dimensional Hilbert space $L^2({\mathbb{R}})$. The quantumness of the system is defined by the Weyl-Heisenberg algebra (\ref{Weyl}) generating $F_{\infty}$. In standard quantum mechanics we identify the generator of $U$ as the {\it momentum} operator, let us say $p$ and the generator of $V$ as the {\it position} operator, let us say $q$ with (\ref{Weyl}) defining the ccr $[p,q]=-i\hbar$. 

Note that when we interpret $F_{\infty}$ as the observer-clock algebra we identify ${\mathbb{R}}$ with "time". In such interpretation $p$ represents, formally (i.e. non positive), the clock "Hamiltonian" and $q$ the {\it clock position} i.e. the marker used to say time. 

What will be relevant for our future discussion on Cosmology is to notice that the {\it dual automorphisms are generated by the {\it clock position} i.e. by the marker used to say time}. We will come back to this discussion in a moment.

\vspace{0.5cm}

Now and before ending this section let us just discuss the definition of traces. Note that the KMS condition for $\omega$ implies that $\omega(\hat a)$ has {\it the trace property}, namely that $\omega(\hat a \hat b)=\omega(\hat b \hat a)$. This is the trace naturally defined on the gauge invariant algebra $(A_{dS})_{\omega}$ and this is the case for {\it any integrable weight $\omega$}. Thus we can define
\begin{equation}
Tr_{\omega}(\hat a)= \omega (\hat a)
\end{equation}
Now since for $A_{dS}$ any integrable weight is {\it dominant} i.e $\omega=\phi\otimes \bar\omega$ we can define the {\it dual automorphism} $\theta_{\phi}^s$ and to identify easily how this dual automorphism changes the value of the trace defined by $Tr_{\omega}$. We get, directly from Weyl relation, the transformation law
\begin{equation}
Tr_{\omega}(\theta_{\phi}^s(\hat a))=e^{-s}Tr_{\omega}(\hat a)
\end{equation}
In essence the {\it dual} automorphism maps the algebra of {\it gauge invariant observables} defined by an integrable weight $\omega$ into the algebra of gauge invariant observables defined by the {\it scaled} integrable weight 
\begin{equation}
\lambda \omega
\end{equation}
with $\lambda=e^{-s}$. 

This simple dual action defines, in the particular case of type $III_1$ factors as $A_{dS}$, the {\it flow of weights}. For factors that are type $III$ but not $III_1$ the flow can be less trivial, however physically we need to restrict our attention to type $III_1$.

The representation (\ref{dual}) of the dual automorphism is a consequence of the known as Takesaki duality \cite{Takesaki}. We can use the dual automorphism $\theta_s$ acting on $(A_{dS})_{\omega}$ to define the crossed product
\begin{equation}
(A_{dS})_{\omega}\rtimes_{\theta_{\phi}^s} {\mathbb{R}}
\end{equation}
The representation (\ref{dual}) of $\theta_{\phi}^s$ implies the isomorphism
\begin{equation}
(A_{dS})_{\omega}\rtimes_{\theta_{\phi}^s} {\mathbb{R}}=A_{dS}\otimes F_{\infty}
\end{equation}
that is the the celebrated Takesaki duality.

\subsection{Type $II_1$ and dual automorphisms}
In \cite{Witten2} it was suggested that empty pure de Sitter is described by a type $II_1$ factor. The physics intuition is that empty de Sitter is the final state describing the static patch once everything inside has gone out the patch and that this state should define a {\it maximal entropy state}. For type $II_1$ factors this state is given by the weight defining the finite and {\it unique} type $II_1$ trace. 

The standard approach to this type $II_1$ factor is to start, as we have just described, with an integrable weight $\omega$ on $A_{dS}$ and to define the gauge invariant algebra $(A_{dS})_{\omega}$ as the corresponding centralizer. This as discussed is a type $II_{\infty}$ factor equipped with
the trace $Tr_{\omega}$ defined in previous section. The type $II_1$ factors are now defined as $p(A_{dS})_{\omega}p$ for $p$ a {\it projection in $(A_{dS})_{\omega}$ with finite value of trace $Tr_{\omega}(p)$}. This value is in essence the way we encode the Gibbons-Hawking entropy of the corresponding empty de Sitter. The maximal entropy state is, as stressed, the weight on $p(A_{dS})_{\omega}p$ defining the finite trace. Note that we have a priori different exemplifications of {\it empty de Sitter} as defined by different i.e. with different values of the trace, finite projections.

The physics way to motivate this definition of the type $II_1$ factor associated to empty de Sitter is to constraint the spectrum of the clock hamiltonian i.e. the generator of the time translations $U^{it}$ entering into the definition of the integrable weight $\omega$, to be {\it positive}. It is now easy to see that:

\vspace{0.3cm}

{\bf Lemma 1:  The type $II_1$ factors $p(A_{dS})_{\omega}p$ are not invariant under the action of the dual automorphisms.}

\vspace{0.3cm}

In practice this means that under the action of the {\it dual automorphisms} $\theta_{\phi}^s$ we can {\it flow} from a type $II_1$ factor defined by a projection $p$ of given finite value $Tr_{\omega}(p)$ into a different type $II_1$ factor defined by a different projection $p'$ with $Tr_{\omega}(p')=e^{-s}Tr_{\omega}(p)$. Thus, the dual automorphism defines flows between examples of {\it empty de Sitter with different GH entropy}.

In more precise terms the {\it flow of weights} defined by the {\it dual automorphisms} induces an associated flow of maximal entropy states. 

The reader could wonder if it will be possible to get a direct path defining a type $II_1$ gauge invariant algebra without going through the type $II_{\infty}$ intermediate step. We will argue that such possibility {\it does not exist}. Indeed if you could define on $A_{dS}$ a weight ,let us say $\tilde \omega$, with centralizer a type $II_1$ factor, then in the Connes-Takesaki ordering of weights we will have that $\tilde \omega \prec \omega$ and consequently $\tilde \omega$ will be integrable (Lemma 2.4 of \cite{CT}) and therefore dominant (Corollary 3.2 of \cite{CT}) making the corresponding centralizer a type $II_{\infty}$ factor. In summary we could conclude

\vspace{0.3cm}

{\bf Lamma 2: There is not any integrable weight $\tilde \omega$ such that $(A_{dS})_{\tilde \omega} = p(A_{dS})_{\omega})p$ for $\omega$ an integrable weight and $p$ a finite projection in $(A_{dS})_{\omega}$.}

\vspace{0.3cm} 

\section{Inflationary Cosmology and the flow of integrable weights}
\subsection{Gauge invariant observables in Inflationary Cosmology}
Let us start recalling some basic expressions defining gauge invariant {\it scalar} fluctuations in a {\it quasi de Sitter inflationary background}. We will follow the original work of Mukhanov and Chibisov \cite{MCH1},\cite{MCH2}. For a technical review, extra references and notation see \cite{MSteinhardt}.

The inflationary background will be characterized by the time dependent metric scale factor $a$ and by the slow roll parameter $\epsilon$ defined by the equation of state
\begin{equation}
\epsilon=\frac{3}{2}\frac{p+\rho}{\rho}
\end{equation}
for $p=\frac{1}{2}{\dot\varphi}^{2}-V(\varphi)$, $\rho=\frac{1}{2}{\dot\varphi}^{2}+V(\varphi)$ and $V(\varphi)$ the inflaton potential. The Hubble constant is given by $H=\frac{\dot a}{a}$ for dot representing derivative with respect to physical time $t=\int a d\tau$ and $\tau$ conformal time.

The metric describing a scalar perturbation will be characterized by
\begin{equation}
ds^2=a^2(\tau)((1+2\phi)d\tau^{2}-(1-2\psi)d{\bar x}^2)
\end{equation}
Given a fluctuation $\delta\varphi$ of the {\it inflaton field} the {\it gauge invariant scalar} fluctuation is given by Mukhanov's variable $v$ defined, during the inflationary period, as
\begin{equation}
v=
a\delta\varphi +z\psi
\end{equation}
with $z=a\sqrt{\epsilon}$. A convenient parametrization in terms of which we can evaluate the power spectrum that we will compare with the CMB experiment is:
\begin{equation}\label{fluctuation}
\chi=\frac{v}{z}=\frac{\delta\varphi}{\sqrt{\epsilon}} +\psi
\end{equation}
The last expression (\ref{fluctuation}) already illustrates the problem of defining {\it gauge invariant} variables in the {\it pure de Sitter limit} $\epsilon=0$. Indeed in this limit the gauge invariant variable is ill defined i.e. divergent in the pure de Sitter limit. Following our former discussion in \cite{Gomez1,Gomez2} we will address this problem from the algebraic point of view. 

\subsection{Back to the definition of the algebra of gauge invariant observables}
In previous sections we have observed that for $A_{dS}$ a type $III_1$ factor and for $\phi$ any {\it state} on $A_{dS}$ the gauge invariant observable defined as
\begin{equation}\label{gauge}
\int_{-\infty}^{+\infty} dt \sigma_{\phi}^t(a)
\end{equation}
for $a$ any local observable in $A_{dS}$ is {\it divergent}. The suggestion in \cite{Gomez1,Gomez2} was {\it to identify} the divergence of (\ref{gauge}) with the divergence of a generic gauge invariant scalar fluctuation (\ref{fluctuation}) in the pure de Sitter limit $\epsilon=0$. This observation has been recently stressed, in a different version, in \cite{Penington,K}. In order to clarify the differences and similarities it could be  worth to design a unified point of view in the language of {\it integrable weights} discussed in previous sections.

In the language of {\it integrable weights} we should expect {\it the existence of a cosmological integrable weight} $\omega_{cos}$ defined on $A_{dS}$ such that
\begin{itemize}
\item The gauge invariant fluctuation $\chi$ defined in (\ref{fluctuation}) for some $\epsilon\neq0$ should be an element of the gauge invariant algebra defined by the weight $\omega_{cos}$ i.e. an element of $(A_{dS})_{\omega_{cos}}$.
\item Consequently $\chi$ could be represented as
\begin{equation}\label{condition}
\chi=\int_{-\infty}^{+\infty}dt \sigma_{\omega_{cos}}^{t} (a)
\end{equation}
 for $a$ in $A_{dS}$ representing the scalar fluctuation $\delta\varphi$.
\end{itemize}
An obvious problem of the former general map between Mukhanov's gauge invariant cosmological variables and elements in $(A_{dS})_{\omega_{cos}}$ for some cosmological integrable weight is the cosmological time dependence of the gauge invariant cosmological observables as $\chi$.

In Inflationary Cosmology we expect a non trivial time dependence reflecting the cosmological time evolution of the Universe. We normally model this time dependence on the basis of the evolution of the inflaton expectation value in a given potential $V(\varphi)$, for instance in the slow roll approximation. The question we need to face, in any algebraic description of Inflationary Cosmology, is: 

\vspace{0.3cm}

{\it How can we describe the cosmological evolution, for instance the one explaining the graceful exit, using the algebra of gauge invariant observables $(A_{dS})_{\omega_{cos}}$, for some integrable cosmological weight ?}

\vspace{0.3cm}

The idea  of {\it Cosmology as flow of integrable weights} offers, in our opinion, a natural answer to this question.

Before going into a discussion of the flow of integrable weights it will be worth in order to clarify the notation to relate the rather abstract former description with the more explicit definition of integrable weights in the case of some simple inflaton potentials presented in \cite{Penington}. The way to define the integrable weight is in terms of the Hartle Hawking wave function defined by a Euclidean action $S(\varphi)$ for the inflaton. As any integrable weight it contains the piece associated to the weight on $F_{\infty}$ defined by $\bar\omega$. Recall that this piece is fully determined by the generator of translations acting on the Hilbert space $L^2({\mathbb{R}})$ used to represent $F_{\infty}$. This Hamiltonian is defined, in the present case, by the Euclidean action for the inflaton with potential $V(\varphi)=V' \varphi$. This generates the $V'$ term in the integrable weight used in \cite{Penington}. 

\subsection{Cosmology as flow of weights}
Gauge invariant observables in $(A_{dS})_{\omega_{cos}}$ for any integrable weight are, by definition, {\it invariant under modular time reparametrizations}. Thus a natural definition of {\it cosmological evolution}, at least in the inflationary context, will be to identify this evolution with a {\it flow of integrable weights} on $A_{dS}$. More precisely with a one parameter family of weights on $A_{dS}$, let us say $\omega_{cos}(s)$ with $s$ parametrizing the flow and consequently, in this approach, the cosmological time or some appropriated function of the physical cosmological time \footnote{The reference to {\it Inflationary Cosmology} is based on using the $A_{dS}$ type $III_1$ factor.}. 

Once fixed some {\it initial conditions}, let us say $\omega_{cos}(s_0)$, a physical observable $\hat a_0$ i.e. gauge invariant with respect to modular time reparametrizations,  in $(A_{dS})_{\omega_{cos}(s_0)}$, will flow, in the {\it cosmological time}, as
\begin{equation}\label{basic}
\hat a(s)= \theta_{cos}^s(\hat a_0)
\end{equation}
for $\theta_{cos}^s$ the generator of the {\it dual automorphisms} already discussed in previous sections. Recall that this dual automorphism defines the action dual to the modular automorphisms on $(A_{dS})_{\omega_{cos}}$ that for $\omega_{cos}$ integrable weight is the crossed product (\ref{crossed}).

The simple expression (\ref{basic}) is, as discussed in previous sections, a consequence of {\it the simplicity of the flow of weights for $A_{dS}$ a type $III_1$ factor}. In this case the flow of weights reduces essentially to pure weight {\it re-scalings}.

Thus the cosmological time dependence of the observable quantities that we effectively measure is given by
\begin{equation}\label{dependence}
\omega_{cos}(\theta_{cos}^s(\hat a_0)) = e^{-s}\omega_{cos}(\hat a_0)
\end{equation}
So choosing for $\hat a_{0}$ a gauge invariant variable the so defined $s$-dependence defines the cosmological time dependence of the corresponding Mukhanov's variable. As a simple example we can define the flow for $\hat a_{0}$ the inflaton field $\varphi$ and to define, using (\ref{dependence}) the corresponding {\it cosmological time dependent} $\varphi(s)$. 

In this context {\it the Einstein equations} will appear {\it as the consequence of the flow of integrable weights on $A_{dS}$}. This natural connection 
 
 \vspace{0.3 cm}
 
{\it Flow of weights $\Longleftrightarrow$ Einstein equations } 

\vspace{0.3 cm}

is reminiscent of the basic relation in String theory between Einstein equations and renormalization group equations. 

Obviously in order to get any concrete expressions we need to figure out how to relate the abstract flow time $s$ and the physical cosmological time $t$. Once this is done in terms of some concrete $s(t)$ we can see how the flow dependence on $s$ defined by (\ref{dependence}) is related to Einstein equations for instance for the simplest slow roll relation $\dot \varphi ^2\sim \dot H$ for $H$ the Hubble parameter. In the present note we will not address this subtle issue in any detail. However we can make some very general remarks for the simple flow of weights in the case of $A_{dS}$ a type $III_1$ factor.

The key element defining the $s$ dependence in this simple case is contained in the factor $e^{-s}$ in (\ref{dependence}). Recall now the deep {\it quantum meaning} of this factor. This factor defining the {\it dual action} $\theta_{cos}^s$ is a direct consequence of {\it the quantum Weyl relations defining the factor $F_{\infty}$} defining the dominant weight $\omega_{cos}$. Recall that any integrable weight is, in the case of $A_{dS}$, a {\it dominant weight}. Actually in a language motivated by the {\it observer} interpretation of $F_{\infty}$ \cite{Witten2} the generator of the dual $\theta_{cos}^s$ ( for any integrable $\omega_{cos}$ ) can be identified (formally) with the {\it position operator} of the {\it observer-clock} \footnote{This reflects the cosmological intuition that the observer clock position i.e. what we use to say time, is part of the own dynamics of the system, namely the {\it inflaton marker}.} . In this sense the cosmological (flow) dependence $e^{-s}$ {\it cannot be trivialized} while we use $F_{\infty}$ generated by the {\it quantum} Weyl-Heisenberg algebra with generators $U_t$ and $V_s$. We can summarize this discussion in the following Lemma.

\vspace {0.5 cm}

{\bf Lemma 3: Non trivial cosmological evolution (in inflationary language graceful exit) is quantum mechanically mandatory.}

\vspace{0.5 cm}

The reason as said is that vanishing flow ( corresponding to $\epsilon=0$ ) will imply that the Weyl algebra defining $F_{\infty}$ is {\it commutative i.e. $\hbar=0$} which is algebraically impossible.
 \subsection{Is the pure de Sitter limit $\epsilon=0$ algebraically consistent?}
 The previous lemma seems to indicate tat the answer is negative. However the definition in \cite{Witten2} of the gauge invariant algebra of observables for pure de Sitter as the type $II_1$ factor seems superficially to indicate the opposite.
 
  {\it How to understand this conflict ?}
  
  Recall that the type $II_1$ factor associated to de Sitter is given by $p(A_{dS})_{\omega_{cos}}p$ for $p$ a finite projection in $(A_{dS})_{\omega_{cos}}$. The problem of defining {\it gauge invariant observables in the $\epsilon=0$ limit} \footnote{Incidentally the problem of defining gauge invariant variables in the limit $\epsilon=0$ is, to the best of my knowledge, a problem that is never touched in standard reviews of inflationary Cosmology, for instance in the definition of the power spectrum of scalar fluctuations.}
is the problem of discovering a weight let us say $\tilde \omega$ such that 
\begin{equation}
 ( A_{dS})_{\tilde \omega} =p(A_{dS})_{\omega_{cos}} p
 \end{equation}
 for $p$ a finite projection. In other words is the problem of finding a weight such that the corresponding centralizer is a type $II_1$ factor. In {\bf Lemma 2} we have argued that this is impossible for type $III_1$ factors i.e. for the de Sitter algebra $A_{dS}$. Thus the type $II_1$ factor defined by some projection $p$ is {\it not in correspondence with any $\epsilon=0$ limit.}
 
 In that sense the type $II_1$ algebra associated to pure de Sitter is just formal since it does not correspond to any $\epsilon=0$ limit. In other words is a definition of de Sitter algebra that ignores the existence of the dual automorphisms which are not an optional tool but something necessarily existing due to the {\it crossed product representation} of $(A_{dS})_{\omega}$ for any integrable weight $\omega$. In summary we conclude that:

 \vspace{0.3cm}
 
 {\it Pure de Sitter defined as $\epsilon=0$ is algebraically inconsistent.}
 
 \vspace{0.3cm}

The existence of the {\it dual automorphisms} implies that we cannot {\it freeze} their action, on the gauge invariant algebra of observables, imposing {\it eternity} through the condition $\epsilon=0$. 
\section{Acknowledgements}
This work is supported through the grants CEX2020-001007-S and PID2021-123017NB-I00, funded by MCIN/AEI/10.13039/501100011033 and by ERDF A way of making Europe.

\end{document}